\documentclass[11pt,a4paper]{article}
\usepackage{latexsym,amsfonts,graphicx}

\title{ Boundary condition at the  junction}
\author{Mark Harmer, Boris Pavlov, Adil Yafyasov}
\begin{document}

\maketitle

\section{Introduction}
 Practical calculation of transport properties of
quantum networks is often reduced to the  scattering problem for a
one-dimensional differential operator on a  quantum graph, see for
instance \cite{Datt,MMK,NMT,SGB,SGZ}. Quantum graph plays a role
of a solvable model for a two-dimensional network, see
\cite{Kuch02,Kost:Sch1,Kost:Sch2}. Basic detail of the model is a
star-shape element with  a self-adjoint boundary condition at the
node. It was  commonly expected  that the realistic  boundary
condition is defined by the angles between the wires at the node.
For instance the boundary conditions for the T-junction,
\cite{Datt}, is presented in terms of  limit values of the
wave-function on the wires $\left\{ \psi_i \right\}^{3}_{i=1}:=
\vec{\psi}$ and the values of the corresponding outward derivative
$\left\{ \psi^{\prime}_i \right\}^{3}_{i=1}:= \vec{\psi'}$ at the
node:
\begin{equation}\label{bndcnd}
\beta^{-1} \psi_1 = \psi_2 = \psi_3 \, , \quad
\beta\psi^{\prime}_{1} + \psi^{\prime}_{2} + \psi^{\prime}_{3} =
0,
\end{equation}
or in the form
\begin{equation}\label{bndcnd2}
P_0^{\perp}\vec{\psi}=0 \, , \quad P_0\vec{\psi}^{\prime}=0
\end{equation}
with the projection
$$
P_0 = \frac{1}{\beta^2 + 2} \left( \begin{array}{ccc}
\beta^2 & \beta & \beta \\
\beta & 1 & 1 \\
\beta & 1 & 1
\end{array} \right).
$$
The scattering matrix of such a junction is $S =  I - 2 P_0 $, see
\cite{Datt, SGB, Tan:But}. In \cite{Datt} $\beta$ is interpreted
as a  free parameter describing the strength of coupling between
the leg and  the bar of the  T-junction. In \cite{Har6} the
condition ( \ref{bndcnd2}) is used for analysis of spin-dependent
transmission across the quantum ring.

We  derive  the boundary condition (\ref{bndcnd2}) and similar
condition  for  any junction   from \cite{BMPPY} and interpret the
corresponding free parameters $\beta, P_0$.
\section{Intermediate Hamiltonian}
Consider one body scattering problem  on the junction $\Omega$
formed by few 2-d equivalent  semi-infinite wires $\Omega_{j},\,\,
s=1,2,\dots $, attached to the quantum well $\Omega_{int}$ via the
orthogonal bottom sections $\Gamma_{s}$. The corresponding
one-body Hamiltonian for the spin-less electron is  scaled, via
replacement of  energy  $E$ by  the spectral parameter $\lambda =
\hbar^{-2} 2m*E $, to the  standard Schr\"{o}dinger operator with
zero  conditions on the boundary $\partial \Omega$, and a constant
potential in the  wires $q|_{\Omega_s} = q_{\infty}$:
\begin{equation}
\label{Schredinger} {\cal L}\psi = -\bigtriangleup \psi + q \psi.
\end{equation}
We  assume, following \cite{BMPPY}, that the potential on the well
is defined by  the scaled constant electric field $\cal{E}$:
$q|_{\Omega_{int}} = q_{int}(x) = \langle \cal{E},x\rangle , x \in
\Omega_{int}$. The role of the non-perturbed  Hamiltonian is
played by the Schr\"{o}dinger operator ${\cal L}_{out}$ on the
wires with zero boundary conditions on the union $ \Gamma = \cup_s
\Gamma_s$ of the bottom sections, which play roles of solid walls,
separating the well from the  wires:
\begin{equation}
\label{zero_bcond} \psi|_{\Gamma} = 0.
\end{equation}
 The  eigenfunctions of ${\cal L}_{out}$ in the  wires $\Omega_s
: 0< x_s = x <\infty,\, 0< y_s = y <\delta $ are combined of {\it
running waves}
\[
\exp (\pm i \sqrt{\lambda - \pi^2 l^2 \delta^{-2}- q_{\infty}
}x)\, \sqrt{2/\delta} \sin \pi l y/\delta \]
\[:= \exp (\pm i {p_l}x)\,\,\,
e_s^l.
\]
Hereafter we  use on the  wires the local coordinates $x,y,\,\,
0<x,\infty, \, 0< y < \delta $. The eigenfunctions of the operator
$L_{int}$ defined by (\ref{Schredinger},\ref{zero_bcond}) on the
quantum well $\Omega_{int}$ are standing waves. Replacement of the
solid wall condition (\ref{zero_bcond}) by the matching condition
is  a strong perturbation blending the standing waves on the
quantum well with the running waves in the wires. This is a
perturbation on the  continuous spectrum, so the convergence of
the corresponding  series can't be estimated in spectral terms of
self-adjoint operators. In \cite {BMPPY} we  suggested a modified
analytic perturbation procedure based on introduction of an
Intermediate Hamiltonian obtained via appropriate splitting, see
\cite{Glazman} of ${\cal L}$.

Assume that the  Fermi level in the  vires lies in the  middle of
the first spectral band $ 2m* E_F \hbar^{-2}:= \lambda_F = \pi^2
\delta^{-2} 5/2 +q_{\infty}$. In that case all branches of the
continuous spectrum with thresholds $\pi^2 l^2\delta^{-2} +
q_{\infty},\, l\geq 2,$ are closed, that is all exponential
solutions $\exp \pm i {p_{l}}x\,\,\, e_s^l,\, ,\, l\geq 2, $ of
the homogeneous equation $-\bigtriangleup \psi + q \psi = \lambda
\psi$ in the  wires are exponentially decreasing. Impose,
additionally  to (\ref{bndcnd}) the semi-transparent boundary
condition on the bottom section
\begin{equation}
\label{semi_trans} u |_{\Gamma} \bot E_+,\,\, E_+ =
\bigvee_{s=1}^3 e_s^1,
\end{equation}
This  boundary condition prevents excitations from the first
channel  ${\cal H}_+ = E_+ \times L_2(R_+)$ in the  wires from
entering into the quantum well, and,
 vice versa, exiting from the quantum well into the wires,
but does not stop the excitations from  the  closed  channels
 ${\cal H}_- = [\bigvee_{l>1} e_s^l ]\times L_2(R_+)$.
The
corresponding operator ${\cal L}^F$ is  split into the orthogonal
sum of the trivial operator in the open first channel ${\cal H}_+$
\[
l^F u = - \frac{d^2 u}{dx^2} + [q_{\infty}+ \pi^2
\delta^{-2}]u,\,\,,
\]
with zero boundary condition on $\Gamma$, and the {\it
intermediate Hamiltonian} $L^F$ in the orthogonal complement $
{\cal H}^F = \left\{ L_2(\Omega) \ominus{\cal H}_+\right\}$. The
spectrum  $\sigma_a (l^F)$ coincides with the first spectral
branch and the continuous spectrum $\sigma_a (l^F)$ of $L^F$ is
$[4\pi^2/\delta^2 + q_{\infty},\,\infty)$.
\begin{figure}[ht]
\includegraphics[width=3in]{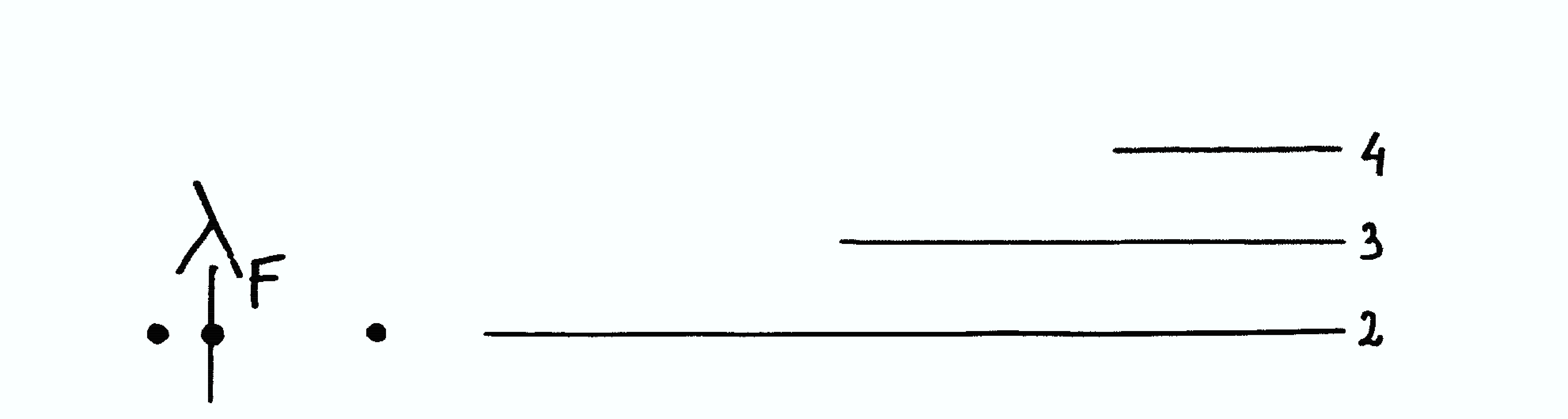}
\caption{Spectrum of  $L^F$}\label{F:figure 2}
\end{figure}
There is a finite number of eigenvalues of $L^F$  situated inside
the first spectral band $\Delta_1= [\pi^2/\delta^2 + q_{\infty},\,
4\pi^2/\delta^2 + q_{\infty}]$ and a countable number of embedded
eigenvalues of $\lambda^F_{r} \in \sigma_a (L^F) $ accumulating at
infinity. 
\section{Scattering matrix via  Intermediate DN map}
Usually the Scattering matrix on the  quantum network is  obtained
via matching  exponential solutions in the wires to the solutions
of the homogeneous Schr\"{o}dinger equation inside the vertex
domain,see, for instance \cite{ML71}. This approach requires
solving an infinite algebraic system. We consider the boundary
problem for the intermediate equation:
 \[
 -\bigtriangleup
\psi + q \psi = \lambda \psi,\,\psi \in L_2 (\omega),
\]
\begin{equation}
\label{interm_bound} \psi|_{\Gamma} = \psi_{\Gamma}\in E_+,
\,\,\Im \lambda \neq 0.
\end{equation}
The  solution $\psi$ exists for all complex $\lambda$  and has
normal limit values  on the continuous spectrum. We introduce the
Dirichlet-to-Neumann map of the intermediate Hamiltonian (DN-map)
as
\begin{equation}
\label{DN_interm} {\cal DN}^F: \psi_{\Gamma} \longrightarrow
P_+\frac{\partial \psi}{\partial n} \psi_{\Gamma},
\end{equation}
with $P_+:= P_{E_+}$. It is a  $3\times 3$ matrix-function which
is obtained via differentiation with respect to  exterior normal
of the resolvent of the intermediate operator restricted onto
$\Omega_{int}$ and framed by $P_+$. It has the kernel:
\[
{\cal DN}^F (y,\eta)= - P_+ \frac{\partial ^2
G_{\lambda}(y,\eta)}{\partial n_y \partial n_\eta} P_+.
\]
It has the spectral representation on the complement of the
spectrum of $ \sigma (L^F)$
\begin{equation}
\label{fewpole} {\cal DN}^{F}(\lambda) =
\sum_{r}\frac{P_+\frac{\partial \phi_{r}(\xi)}{\partial n}\rangle
\,\langle  P_+\frac{\partial \phi_{_{r}}(\xi')}{\partial
n}}{\lambda- \lambda_{r}} + {\cal K}^F(\lambda),
\end{equation}
where the summation is  extended over discrete spectrum of $L^F$
and ${\cal K}^F(\lambda)$ contains an integral over the continuous
spectrum of $L^F$. The  scattering matrix of ${\cal L }$ is
obtained via matching of the  scattering Ansatz  on the open
channel in wires with $p = p_1$:
\begin{equation}
\label{SansatzF} e^{^{i p\xi}} e_{_{+}} + e^{^{- i p\xi}} S(p)
e_{_{+}}
\end{equation}
to the limit values on the spectrum, $\Im \lambda \to 0$ of the
solution of the above intermediate boundary problem
(\ref{DN_interm}):
\[
\label{F_matching} ip\left[ e_{_{+}} - S(p) e_{_{+}}\right] =
{\cal DN}^{F} (\lambda)\, \left[e_{_{+}} +  S(p) e_{_{+}}\right].
\]
Solving this  equation we  obtain, see  \cite{BMPPY}, the formula
for the  scattering matrix  of the operator ${\cal L}$ on the
first spectral band $\Delta = [\pi^2/\delta^2 +
q_{infty},\,4\,\pi^2/\delta^2 + q_{infty},\,]$ in terms of ${\cal
L}^{F}$ by the formula
\begin{equation}
\label{SmatrixF} S(p) = \frac{ {\cal DN}^F (\lambda) + ip P_{+}}
{-{\cal DN}^F (\lambda)+ ip P_{+}}.
\end{equation}
The DN-map $ {\cal DN}^{^{F}} $ of the intermediate Hamiltonian
$L_{_F}$ is connected  with  the standard \cite{SU2} DN-map $
\Lambda $ of the operator  $L_{int}$ on the quantum well $\Omega$
by the formula
\begin{equation}
\label{DNF_DN} {\cal DN}^{^{F}} = P_+ {\cal DN} P_+ - P_+ {\cal
DN}
 P_{_{-}}\,\,{\bf D}^{-1}
     P_- {\cal DN} P_+ .
\end{equation}
Here
\[
 {\bf D} = P_- {\cal DN}P_- - \,K_{_{-}},\,\,\]
with
 \[
 K_- = -\oplus
\sum_{l>1} p_l P_l,\, |K_-| \geq \pi \delta^{-1} \sqrt{3/2}.
\] and
$P_-:= \oplus\sum_{l>1}\sum_{s} e_s^l\rangle \langle e_s^l$. Near
the eigenvalue $\lambda_0$ of $L_{int}$ the DN-map  can be
decomposed as
\begin{equation}
\label{formal_decomp} {\cal DN}^F(\lambda) =
\frac{\phi_{_{0}}\rangle \langle \phi_{_{0}} }{\lambda -
{\lambda_0}}
 + \sum_{s\neq
0} \frac{\phi_{_{s}}\rangle \langle \phi_{_{s}}} {\lambda -
{\lambda_s}} := \frac{\phi_{_{0}}\rangle \langle
\phi_{_{0}}}{\lambda - {\lambda_0}} + {\cal K}_{{0}},
\end{equation}
where  $\phi_{_{s}} = P_{_{-}}\frac{\partial \Phi_{_{s}}}{\partial
n}|_{{\Gamma}},\,s= 0,1,\dots $ are the  boundary currents of
eigenfunctions $\Phi_{_{s}},\, s= 0,1,2\dots$ of the  operator
$L_{int}$, $\lambda_{{s}}$\,- the  corresponding eigenvalues.
Spacing between the eigenvalues of $L_{int}$ is connected  to the
diameter $d_{int}$ of $\Omega_{int}$ as $\rho_0 = \min_{r\neq
0}|\lambda_0 - \lambda_r|= O(d_{int}^{-2}$. Due to the spectral
estimate  $|{\cal K}_{{0}} (\lambda_0)| \approx \rho_0^{-1}$. For
relatively  thin networks the analytic perturbation procedure  can
be  developed based on (\ref{DNF_DN}), with the small parameter
$\delta/d_{int}$, to obtain ${\cal DN}^F$. Denote by ${\cal
K}_{\pm,\pm}$ the matrix elements  $ P_{\pm}{\cal K}_{{0}}P_{\pm}$
with respect to the orthogonal decomposition $L_2 (\Gamma) = E_+
\oplus E_-$. If only one eigenvalue $\lambda_0$ of $L_{int}$  is
situated of the  essential spectral interval $\Delta_T =
[\lambda^F - 2 \hbar^2 m^* \kappa T,\, \lambda^F + 2 \hbar^2 m^*
\kappa T$, we can represent the  denominator $\bf D $ of
(\ref{formal_decomp}) near $\lambda_0$, with  $P_- \phi_0 :=
{\phi}^{{-}}_0$  with a  controllable  error $k$:
\[
 {\bf D}(\lambda) =
\frac{{\phi}^{{-}}_0\rangle \langle {\phi}^{{-}}_0} {{\lambda} -
{\lambda_0}} + {{\cal K}}_{--} - K_- : =
\frac{{\phi}^{{-}}_0\rangle \langle {\phi}^{{-}}_0} {{\lambda} -
{\lambda^0_0}} + k
\]
so that the  whole expression (\ref{DNF_DN}) can be calculated via
analytic perturbation procedure, since $k \approx \delta^{-1} [1 +
d_{int}^{-2} ] >> 1$:
\begin{equation}
\label{DN_approx} \Lambda^{^{F}} \approx \frac{\phi_{_0}^{^F}
\rangle \,\langle \phi_{0}^{^F} }{\lambda - \lambda_0^F},
\end{equation}
with $ \lambda_0^F = {\lambda}_{0}  + \langle P_- \phi_{0},\,
k^{^{-1}}\,\, \, P_- \phi_{0}\rangle \approx \lambda_0 $ and
$\phi_{0}^{F} = P_+\phi_{0} - {\cal K}_{{+ -}}
k^{{-1}}\,\,P_-\phi_{0} \approx P_+\phi_{0} $. For low temperature
only electrons with energy close to Fermi level $E_F$ contribute
to transport phenomena. Hence ${\cal DN}^F$ may be substituted by
the single resonance term $\frac{P_0 {\phi}^-_{{0}}
\rangle\,\langle P_0{\phi}_{_{0}}}{\lambda_{0} - \lambda}, $ thus
resulting in the approximate expression for the scattering matrix
on $ \Delta_{T}$ :
\begin{equation}
\label{Sapprox} S_{approx} (\lambda) = \frac{ip P_0 - \frac{
{P_0\phi}_{{0}} \rangle\,\langle {P_0\phi}_{{0}}}{\lambda_{0} -
\lambda}}{ip P_0 + \frac{ P_0{\phi}_{{0}} \rangle\,\langle
P_0{\phi}_{{0}}}{\lambda_{0} - \lambda}},
\end{equation}
see \cite{BMPPY} and more details in \cite{MPP04}.
\section{Boundary condition at the  junction}
The approximate scattering matrix can be obtained from the
energy-dependent boundary condition at the vertex imposed onto the
scattering Ansatz (\ref{SansatzF}) in the  wires:
\begin{equation}
\label{bcond3} ip [I - S_{approx}(\lambda)]\vec{\psi} = [I +
S_{approx}(\lambda)]\vec{\psi}'.
\end{equation}
The polar terms in the numerator and in the denominator of
(\ref{Sapprox}) have the dimension $cm^{-1}$ and can be
represented via the relevant one-dimensional orthogonal projection
$P_{_{0}}:= \vec{e}_{_{0}}\rangle\,\langle \vec{e}_{_{0}} $ with
$\vec{e}_{_{0}} := (e_0^1,e_0^2,\dots e_0^n)=
\parallel \vec{\phi}_{_{0}} \parallel^{-1} \vec{\phi}_{_{0}} :=
\alpha^{-1} \vec{\phi}_{_{0}}$. Then  $
\vec{\phi}_{_{0}}\rangle\,\langle \vec{\phi}_{_{0}} =
\alpha^{^{2}}P_{_{0}}$.
Denoting by $P^{^{\bot}}_{_{0}}$ the complementary projection $I -
P^{^{\bot}}_{_{0}}$ in $L_{2}(\Gamma)$, we  obtain
\begin{equation}
\label{approxS} S_{approx}(\lambda) = P^{^{\bot}}_{_{0}} +
\left[\frac{ip(\lambda - \lambda_{0}) + \alpha^{^{2}} }{ip(\lambda
- \lambda_{0}) - \alpha^{^{2}}}\right] P_{_{0}}
\end{equation}
\[
 \equiv P^{^{\bot}}_{_{0}} + \Theta(\lambda) \, P_{_{0}}.
\]
The factor $\Theta$ is  close  to $-1$ on the essential spectral
interval $\Delta_T$, for low temperature $ 2\pi m*\sqrt{3/2}
\kappa T << \delta \alpha^2 \hbar^2$. Then, in first
approximation, the energy-dependent boundary condition
(\ref{bcond3}) is reduced on $\Delta_{T}$ to $i P^{{\bot}}_{{0}}
\psi - P_{{0}} \psi'\approx 0 $, or, due to orthogonality of
$P_{{0}},P^{{\bot}}_{{0}}$, to $ P^{{\bot}}_{{0}} \vec{\psi}
\approx 0;\, P_{_{0}} \vec{\psi}' \approx 0$. This condition
 coincides with the  above  Datta condition (\ref{bndcnd}) presented
 in form (\ref{bndcnd2}). Our analysis reveals the meaning of the
 projection $P_0$: it coincides with the projection onto the
 one-dimensional subspace defined by the vector
 $P_+\frac{\partial \phi_{0}(\xi)}{\partial n}|_{\Gamma}$
 of boundary values of the normal derivatives of the  resonance
 eigenfunction, projected onto $E_+$. This  formula is  valid also
 for  general junctions.
\begin{figure}[ht]
\includegraphics[width=3in]{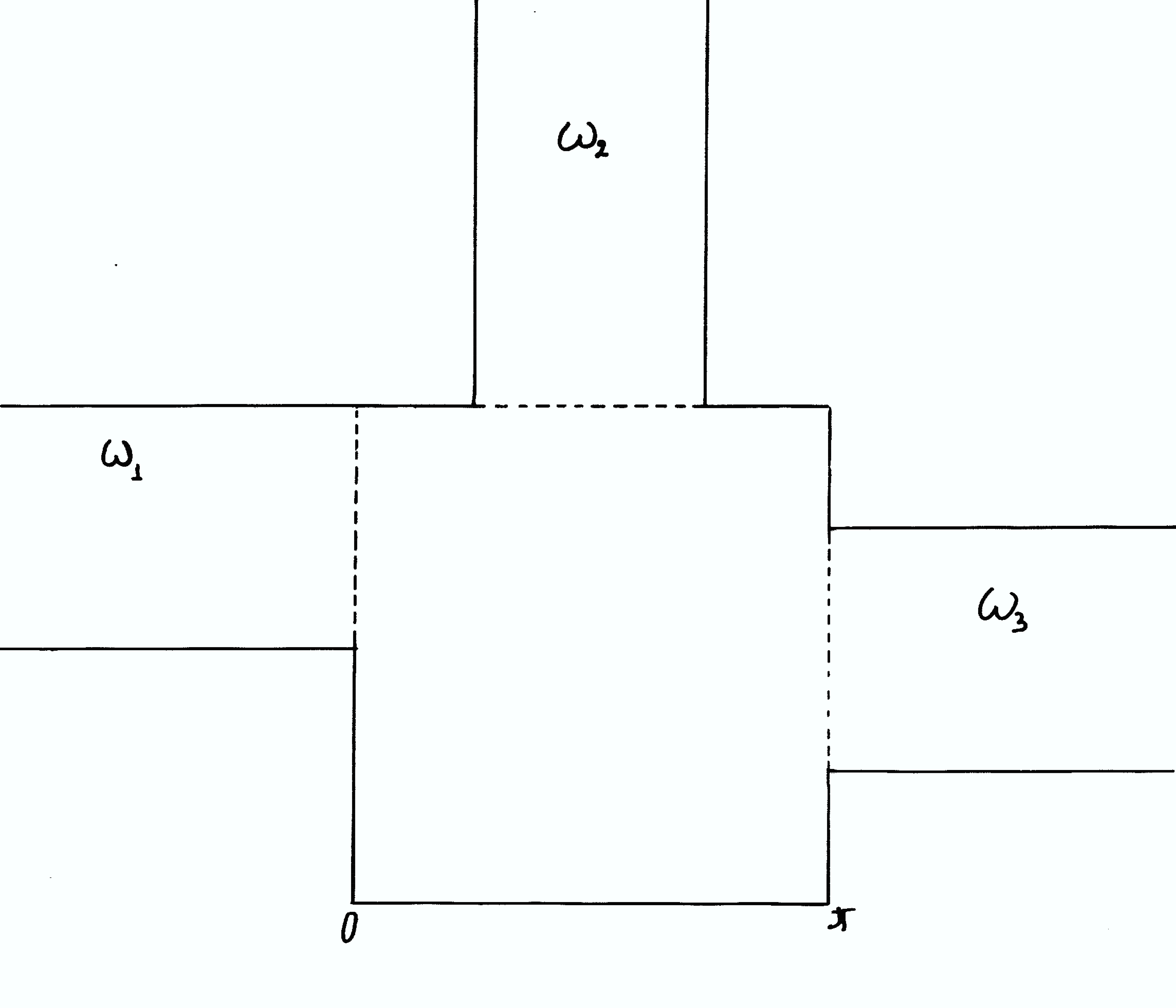}
\caption{ Simplest asymmetric T-junction}\label{F:figure 4}
\end{figure}
\section{Example}
Consider a  two-dimensional quantum network constructed as a
simplest asymmetric  T-junction of three straight quantum  wires
width $\pi/2$  attached to the quantum well - the square
$\Omega_{int}$ on $\xi$-plane : $ 0< \xi_{_{1}}< \pi,\, 0<
\xi_{_{2}}< \pi$. Assume that the first wire $\omega_{{1}}=
\left\{ -\infty< \xi_{_{1}}< 0,\,\pi/2< \xi_{_{2}}< \pi \right\}$
is attached orthogonally to the  left side  of $\partial \Omega$
on $\Gamma_{{1}} = \left\{ \xi_{{1}} = 0,\, \pi/2< \xi_{{2}}< \pi
\right\}$,\, the second wire $\omega_{_{2}} = \left\{ 0<
\xi_{{1}}< \pi/2,\,\pi< \xi_{{2}}< \infty \right\}$ is  attached
in the  middle of the upper side, and the third wire
$\omega_{_{3}}= \left\{ \pi < \xi_{_{1}}< \infty,\, 0< \xi_{_{2}}<
\pi \right\} $ is  attached to the middle of the right side of
$\Omega_{int}$.
On the constructed quantum network $\Omega = \Omega_{int}\cup
\omega_{{1}}\cup \omega_{{2}}\cup\omega_{{3}}$ consider the
scattering problem for Laplacian with homogeneous Dirichlet
boundary condition at the boundary. The cross-section
eigenfunctions in the first channel in the wires
$\omega1,\,\omega-2,\,\omega_3$ are :
\[
e^1 \bigg|_{\Gamma_1} = \frac{2}{\sqrt{\pi}} \sin 2 \xi_2,\,\, e^2
\bigg|_{\Gamma_2} = \frac{2}{\sqrt{\pi}} \cos 2 \xi_1,\,\, \]
 \[
e^3 \bigg|_{\Gamma_3} = \frac{2}{\sqrt{\pi}} \cos 2 \xi_2
\]
The Dirichlet  Laplacian on $\Omega_{int}$ has on the  first
spectral band $\Delta_1 = [4,16] $ the eigenvalues  $
\lambda_{_{0}} = 5 $ ,$ \lambda_{1} = 8 $, $ \lambda_{_{2}} = 10$
and $\lambda_{_{3}} = 13 $ with eigenfunctions
$\Phi_0,\Phi_1,\Phi_2.\Phi_3$, $\Phi_0 = \frac{2}{\pi} \sin \xi_1
\sin 2\xi_2 $. The boundary currents of $\Phi_0$ are
\[
\frac{\partial \Phi_0}{\partial n}\bigg|_{\Gamma_1} =
-\frac{2}{\pi}\sin 2\xi_2,\,\frac{\partial \Phi_0}{\partial
n}\bigg|_{\Gamma_2} = -\frac{4}{\pi}\sin \xi_1,\,
\]
\[
\frac{\partial \Phi_0}{\partial n}\bigg|_{\Gamma_3}=
\frac{2}{\pi}\sin 2\xi_2.
\]
Assume that the Fermi level of the material is situated between
the first and the second thresholds of the network $4 < E_{_{F}}<
16 $ close to the eigenvalue $\lambda_{0} = 5 \,\, cm^{^{-2}},\,
\lambda^F=4.33\,\,cm^{-2}$. The  electrons are supplied to the
network in the first  spectral band  from the  second wire across
the  bottom section $\Gamma_2$ and exit across
$\Gamma_1,\,\Gamma_3$. Due to orthogonality of the cross-section
eigenfunction of the open channel to the boundary currents of the
eigenfunctions $\Phi'_{{0}},\,\Phi'_{{3}} $ the corresponding
modes are not excited. An essential link to the closed channels is
supplied only by  $\Phi_{{0}}$, the contribution from other
eigenfunctions either vanish or are suppressed due  to the factors
$(\lambda_{{0}} - \lambda_{s} )$ in the denominator. The link of
$\Phi_{{0}}$ only to the closed channel in $\omega_{{3}}$ gives a
scalar equation  for the re-normalized eigenvalue
$\hat{\lambda}_{{0}}$ of the  intermediate Hamiltonian, since only
the  contribution from $\Gamma_{_{3}}$
 is  non-trivial:
\[
\frac{1}{\hat{\lambda}_0 - 5}
\int_{_{\Gamma_{_{3}}}}\bigg|P_{_{-}}\frac{\partial
\Phi_{_{0}}}{\partial n}\bigg|^{^{2}} d\Gamma_{_{3}} + \sqrt{16 -
5 } := {\bf D}_{_{-}}(\hat{\lambda}_0) =
\]
\begin{equation}
\label{shift} \sqrt{11}\,\,[0.67 + \hat{\lambda}_0 - 5]
\end{equation}
From this equation  we  obtain the resonance eigenvalue of the
intermediate Hamiltonian: $\hat{\lambda}_0 = 4.33 =
\lambda^{{F}}$. The boundary current of the corresponding
eigenfunction essentially coincides with the boundary current of
the normalized resonance eigenfunction $\Phi_{_{0}}$ of the
Dirichlet Laplacian on $\Omega_{{0}}$. The  projections of the
resonance boundary currents onto $E_+$ are
\[
\phi_0^1 = \int_{{\Gamma_{{1}}}}\, \frac{\partial
\Phi_{{0}}}{\partial n}\bigg|_{{\Gamma_{{1}}}}\, e^{{1}}_{{+}}
\bigg|_{{\Gamma_{{1}}}} d \xi_{_{2}} = - \frac{1}{\sqrt{\pi}} = -
0.56,
\]
\[
\phi_0^2 = \int_{_{\Gamma_{{2}}}}\, \frac{\partial
\Phi_{{0}}}{\partial n}\bigg|_{{\Gamma_{{2}}}}\, e^{{2}}_{_{+}}
\bigg|_{{\Gamma_{{2}}}} d \xi_{{1}} = \frac{16\, \sqrt{2}}{3
\pi\sqrt{\pi}} = 0.43,
\]
\[
\phi_0^3=\int_{{\Gamma_{{3}}}}\, \frac{\partial
\Phi_{{0}}}{\partial n}\bigg|_{{\Gamma_{{3}}}}\, e^{^{3}}_{{+}}
\bigg|_{{\Gamma_{{3}}}} d \xi_{{2}} = 0.
\]
Then the normalized vector of the boundary current is  $e_{{0}}
=\left( - 0.8,\, 0.6,\, 0 \right) $, and the  boundary conditions
at the  junction for low  temperatures are represented by  the
 formulae (\ref{bndcnd2}) with $P_{{0}} = e_{{0}}\rangle\, \langle
e_{_{0}}$, which is  different from  the condition for  a
symmetric junction suggested in \cite{Datt} for  symmetric
T-junction. For the higher temperatures  the boundary condition is
energy dependent and can be represented in form (\ref{bcond3}),
with the approximate scattering matrix
\[
S_{{appr}}(p) = \frac{i \sqrt{\lambda -4}P_{{+}} - 0.15
\frac{P_{{0}}}{\lambda - 4.33}}{i \sqrt{\lambda -4}P_{{+}} + 0.15
\frac{P_{{0}}}{\lambda - 4.33}},
\]
with $P_0 = e_0 \rangle\langle e_0$.

\end{document}